\documentclass[epj]{svjour}
\usepackage{graphics}
\begin{document}
\title{Limits of structure stability of simple liquids revealed by study of relative fluctuations}
\author{A.L. Goncharov\inst{1} \and V.V. Melent'ev\inst{2} \and E.B. Postnikov\inst{1}
}                     
\institute{Department of Theoretical Physics, Kursk State University,
Radishcheva st., 33, 305000, Kursk, Russia \email{postnicov@gmail.com} \and Laboratory of Molecular Acoustics, Kursk State University,
Radishcheva st., 33, 305000, Kursk, Russia}
\date{}
%
\abstract{
We analyse the inverse reduced fluctuations (inverse ratio of relative volume fluctuation to its value in the hypothetical case where the substance acts an ideal gas for the same temperature-volume parameters) for simple liquids from experimental acoustic and thermophysical data along a coexistence line for both liquid and vapour phases. It has been determined that this quantity has a universal  exponential character within the region close to the melting point. This behaviour satisfies the predictions of the mean-field (grand canonical ensemble) lattice fluid model and relates to the constant average structure of a fluid, i.e. redistribution of the free volume complementary to a number of vapour particles. The interconnection between experiment-based fluctuational parameters and self-diffusion characteristics is discussed. These results may suggest experimental methods for determination of  self-diffusion and structural properties of real substances.
} 
\maketitle
\section{Introduction}

One of the trends of modern condensed matter physics is revitalising theories of simple liquids, see e.q. \cite{Bertolini2011,Zhou2011,Ingebrigtsen2012}. First of all, this is connected with the recent emergence of computational opportunities for a  simulation of many-particles systems. Correspondingly, the majority of recent works deal with the detailed analysis of statistical and thermodynamical model systems, such as hard spheres, Lennard--Jones, Morse potentials, etc. \cite{Chakraborty2007,Liu2012,Ingebrigtsen2012}. 

On the other hand, there are questions regarding the limits of their applicability to real fluids and whether the accuracy of their predictions is suitable for technical applications. Worthy of particular note are the lattice gas models pioneered by Fowler \& Guggenheim \cite{Fowler1940} and Lee \& Yang \cite{Lee1952}, and further developed, e.g. by Sanchez \& Lacombe \cite{Sanchez1976} into a form suitable for quantitative description of practically important substances (noble gases, moderate molecular weight hydrocarbons, some polymers). Due to the simplicity of its implementation and clear physico-geometrical structural picture, the last class of models is still under active development and being adjusted to more realistically reproduce thermophysical properties of substances, see the review of the current state of the art in \cite{Binder2011}.

However, the majority of these works deals with a thermodynamical picture, where parameters are adjusted to satisfy  PVT characteristics, and which have relatively small sensitivity to details of microstructure. From this point of view, the study of relative volume fluctuations  should provide much more information about structural characteristics. It especially  concerns the region of high densities, close to the melting point, where the applicability of various theories of liquid is still under discussion.   

The main goal of the present work is to study fluctuation properties in real fluids based on acoustic (speed of sound) and thermophysical (heat capacity ratio) measurements along a coexistence line of liquid and vapour phases. This molecular acoustic approach allows us to extract the required information from data which are i)highly accurate due to the relatively simple experimental techniques and ii) averaged over a macroscopic volume that makes them most suitable for the realisation of the second goal of this work: to test the qualitative patterns provided by the lattice gas model with the real fluctuational behaviour of dense liquids near their melting points.

\section{Inverse relative fluctuation derived from acoustical data}

The relative volume fluctuation in an arbitrary isotropic fluid can be written as \cite{Stanley}
\begin{equation}
\frac{\left\langle (\Delta V)^2\right\rangle}{V}=\beta_Tk_B T,
\label{rel_Vfluct}
\end{equation}
where $\beta_T$, $k_B$ and $T$ are the isothermal compressibility, Boltzmann's constant and temperature, respectively.

For an ideal gas, equation (\ref{rel_Vfluct}) takes the form
\begin{equation}
\frac{\left\langle (\Delta V)^2\right\rangle}{V}=\frac{\mu_0k_B}{\rho R}=\frac{V}{N},
\label{rel_VfluctIG}
\end{equation}
where $\mu_0$, $\rho$, $R$ are the molar mass, density and gas constant, correspondingly; $N$ is the number of particles. 

Thus, dividing (\ref{rel_VfluctIG}) by (\ref{rel_Vfluct}) and using the following expressions for the heat capacity ratio, $\gamma=C_p/C_v=\beta_T/\beta_S$, and speed of sound, $c=1/\sqrt{\beta_S\rho}$, we obtain the parameter
\begin{equation}
\nu=\frac{\mu c^2}{\gamma RT}.
\label{nu}
\end{equation}
In fact, this parameter is a ratio of speeds of sound for the substance and the model ideal gas of the same molecules at the same density and temperature. 
 Thus, the departure of parameter (\ref{nu}) from unity is determined by the strength of intermolecular interactions in a given medium. At the same time, this interaction could take in to account not only long-range forces, but also geometrical packing. This is because expression $S(0)=\beta_Tk_B\rho T$ connects the structure factor $S(q)$ at the wave number $q=0$ with the relative fluctuation of the number of molecules in the scattering volume $V$ \cite{Stanley}.

\begin{figure}
\resizebox{0.5\textwidth}{!}{\includegraphics{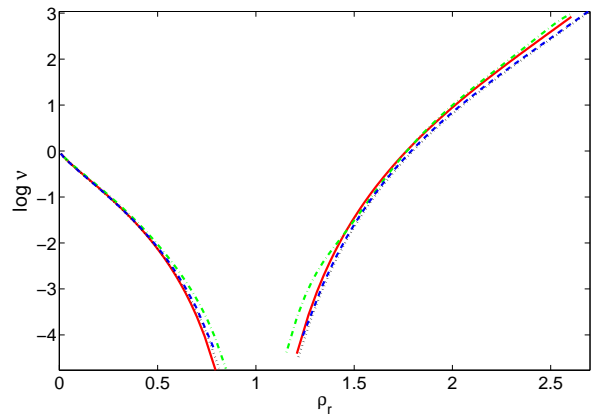}}
\caption{(Colour online) The semilogarithmic plot of the reduced inverse fluctuation as a function of the reduced density for noble gases along a coexistence line. Green (dash-dotted), red (solid), blue (dashed), and black (dotted) curves correspond to neon, argon, krypton, and xenon. Left and right sets of curves are dependencies obtained for vapour and liquid branches, correspondingly.}
\label{ng_fluct}
\end{figure}

To analyse the behaviour of parameter (\ref{nu}) along the coexistence line, we use the data describing the  density, speed of sound and heat capacity ratio (as well as thermodynamical constants) taken from the {\tt  NIST Chemistry WebBook} \cite{NIST}. The resulting curves are presented in Fig.~\ref{ng_fluct} as functions of the reduced density $\rho_r=\rho/\rho_c$, where $\rho_c$ is the critical density. Because of the  corresponding state's law the curves for all substances are close to each other.  Here we plot not the inverse reduced fluctuations but rather their logarithms, since this representation reveals the region of our interest: both sets of curves in Fig.~\ref{ng_fluct} exhibit intervals where they can be fitted linearly with good accuracy. 

\begin{figure}
\resizebox{0.5\textwidth}{!}{\includegraphics{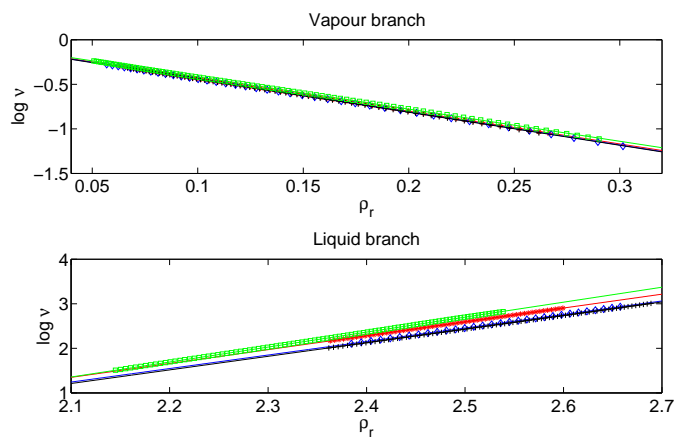}}
\caption{(Colour online) The linear fits (straight lines) of experimental data for neon (green squares), argon (red asterisks), krypton (blue diamonds), and xenon (black dots) within regions of linear fit admissibility. Corresponding equations and approximation accuracies are presented in Table~\ref{tab}.}
\label{ng_deriv_fluct}
\end{figure}

To check this semilog property (in other words, the exponential growth/decay of the inverse fluctuations for a saturated liquid/vapour with the increasing  density), we have presented $\log \nu$ for this density interval separately, see Fig.~\ref{ng_deriv_fluct}. Corresponding linear fit equations and the norm of residuals are listed in the Table~\ref{tab}. The small magnitude  of the residuals confirms the visual quality of this approximation for sufficiently long intervals for all studied liquids (up to $25\%$ of the reduced density range for each phase, compare scales in Fig.~\ref{ng_fluct} and Fig.~\ref{ng_deriv_fluct}). 
In addition, the values of the slopes are relatively close  for both vapour and liquid branches. The possible origin of this similarity will be discussed below.

\begin{table}
\caption{Linear fit equations for experimental data and lattice model under saturation conditions.}
\label{tab}       
\begin{tabular}{ccc}
\hline\noalign{\smallskip}
Substance & Fit & Norm of residuals  \\
\noalign{\smallskip}\hline\noalign{\smallskip}
&Liquid branch&\\
\noalign{\smallskip}\hline\noalign{\smallskip}
Ne & $3.358\rho_r-5.695$ & $0.015$ \\
Ar & $3.111\rho_r-5.184$ & $0.003$  \\
Kr & $3.033\rho_r-5.126$ & $0.002$  \\
Xe & $3.050\rho_r-5.194$ & $0.004$  \\
Lattice & $3.454\rho_r^*-10.4$ & $0.024$  \\
\noalign{\smallskip}\hline\noalign{\smallskip}
&Vapour branch&\\
\noalign{\smallskip}\hline\noalign{\smallskip}
Ne & $-3.610\rho_r-0.056$ & $0.020$ \\
Ar & $-3.680\rho_r-0.065$ & $0.013$  \\
Kr & $-3.706\rho_r-0.069$ & $0.015$  \\
Xe & $-3.709\rho_r-0.0713$ & $0.010$  \\
Lattice & $-2.637\rho_r-0.057$ & $0.009$  \\
\noalign{\smallskip}\hline
\end{tabular}
\end{table}

To discuss the origin of such exponential behaviour, let us note that an expression similar to (\ref{rel_Vfluct}),
$
D_{id}/D=\beta_Tk_B\rho T,
$
is known \cite{Skripov} as the ratio of the self-diffusion coefficient of a dense medium having the properties of an ideal gas ($D_{id}$) to an actual self-diffusion coefficient. At the same time, a universal scaling law has been phenomenologically proposed  for atomic diffusion in condensed matter \cite{Dzugutov1996} which reads $D^*\sim\exp(S_2)$, where $D^*=D/D_{HS}$ is a dimensionless  self-diffusion coefficient ($D_{HS}$ marks the coefficient for the corresponding hard-sphere system), and 
\begin{equation}
S_2=-2\pi\rho\int_0^{\infty}\{g(r)\log[g(r)]-[g(r)-1]\}r^2dr
\label{entropy}
\end{equation}
is the two-particle approximation of the excess entropy (the difference between the actual entropy and the entropy of the equivalent ideal gas) represented as a function of the radial distribution function $g(r)$.

It has been argued that the  excess entropy extracted from self-diffusion data  is an important and sensitive parameter for the study of  structural transitions in liquids, see the work \cite{Errington2006} where the method was first proposed and \cite{Yan2008} where it was especially studied in particular detail. Thus, this exponential dependence of the compressibility on the density suggests that the excess entropy (and thus the structure) is independent of the density, as follows from the aforementioned relations for the reduced self-diffusion coefficient and the formula (\ref{entropy}).

Since usually a fluid structure is determined by the first coordination shell, and $g(r)$ is close to 1 both for liquids and gases \cite{Touba1997}, Eq.~(\ref{entropy}) can be represented \cite{Yan2008} as 
$$
S_2\approx-2\pi\rho\int_0^{\infty}\left[g(r)-1\right]^2r^2dr=-2\pi\rho\int_0^{\infty}|h(r)|^2r^2dr,
$$
where $h(r)$ is a total spatial pair correlation function. 

Using the asymptotic behaviour $h(r)\sim r^{-1}\exp(-r/\xi)$, the excess entropy finally reads $S_2\sim-\pi\xi\rho$ with the correlation length $\xi$ as a parameter.

This exponential (linear in semilogarithmic co-ordinates) behaviour, which is shown in Fig.~\ref{ng_deriv_fluct}, indicates a constant correlation length for the extracted intervals, i.e. an absence of structural transitions there. For liquid branches, these intervals start from the melting point. On the other hand, the right boundary of the interval marks the density, where the correlation length starts to grow (up to infinity at a critical point). Thus, the representation of measurements such as Fig.~\ref{ng_fluct} allows the subdivision of  density intervals into a subinterval of normal behaviour and an interval of a scale invariance closed to a critical point for real, not model, systems.  

The set of vapour branches also exhibits a region of linearity in the semilogarithmic scale.  This confirms the theoretical prediction \cite{Rosenfeld1999} that the Dzhugotov's exponential universality  also holds for dilute gases. In addition, the slopes  of the linear fit lines for the vapour branches are close to the slopes for the corresponding liquids within one temperature range, but have opposite signs (some difference in the slopes corresponds to more dense packing of particles in the liquid due to attractive interactions). This supports the applicability of the hole theory of liquids \cite{Skripov} for these systems. The aforementioned behaviour describes the spatial distribution of vapour molecules as if they fill the holes in the coexisting liquid.

To validate such a model of fluctuational behaviour,  in the next section we consider  the simple lattice fluid model with a variable number of particles along a coexistence curve. Such a model  deals directly with a ``fluid with holes'' representation.

\section{Fluctuations in a saturated lattice fluid}

Following \cite{Cervera2011} let us consider the simple lattice fluid model, where the full volume $V_s$ available for particles is subdivided into $N_s=V/b$ cells of elementary volume $b$. Every cell can contain not more than one particle, and nearest-neighbour particles interact with  local attractive energy $w=\mathrm{const}<0$. For the consideration of phase transitions and coexistent states, the mean-field  distribution function is expressed as a function of temperature $T$, full number of cells (maximal value of discrete volume variable) $N_s$ and the chemical potential $\mu$ as follows
$$
P(N)=\frac{1}{\Xi(T,N_s,\mu)}\left({N_s \atop N}\right)\lambda^N\sigma^{\frac{N^2}{N_s}},
$$ 
where $\lambda=\exp(\mu/k_BT)$, $\sigma=\exp(-Z_1w/k_BT)$ (here $Z_1$ is a first co-ordination number), and the  
grand canonical partition 

$$
\Xi(T,N_s,\mu)=\sum\limits_{N=0}^{N_s}\left({N_s \atop N}\right)\lambda^N\sigma^{\frac{N^2}{N_s}}.
$$

Along the coexistence line, using the Fowler-Guggenheim expression for the mean-field chemical potential on a lattice, as a function of temperature $T$ and volume per particle $v=V/N=bN_s/N$,
$$
\mu=\frac{bZ_1w}{v}+k_BT\ln\frac{b}{v-b},
$$
the distribution function takes the form
\begin{equation}
P(N)=\frac{1}{\Xi(T_r,N_s,\mu)}\left({N_s \atop N}\right)e^{\frac{2N}{T_r}\left(\frac{N}{N_s}-1\right)},
\label{distr_l}
\end{equation}
where $T_r$ is a reduced temperature determined as the ratio $T_r=T/T_c$ with critical temperature $T_c=-Z_1w/4k_B$.

The distribution function (\ref{distr_l}) has bimodal character for liquid-vapour coexistence conditions. The critical temperature corresponds to $\langle N \rangle_c=N_s/2$ and merging the two peaks into a unimodal distribution.

Since the reduced averaged squared fluctuation of the volume per constant number of particles and  an averaged squared fluctuation of the number of particles in a constant volume are connected as $\langle (\Delta V)^2\rangle/V^2=\langle\Delta N)^2\rangle/\langle N \rangle^2$, i.e. $\langle (\Delta V)^2\rangle/V=(V/N)\langle(\Delta N)^2\rangle)/\langle N \rangle$, then taking into account Eq.~(\ref{rel_VfluctIG}), we obtain the desired relative fluctuation in the form
\begin{equation}
\nu_{latt}^{-1}=\frac{\langle N^2 \rangle-\langle N\rangle^2}{\langle N\rangle}.
\label{nulatt}
\end{equation}

Corresponding averaged quantities are determined separately as 
\begin{equation}
\langle N\rangle=\sum\limits_{N=N_{min}}^{N_{max}}NP(N), \quad \langle N^2\rangle=\sum\limits_{N=N_{min}}^{N_{max}}N^2P(N),
\label{mN}
\end{equation}
where $\{N_{min},N_{max}\}$ are $\{0,N_s/2\}$ and $\{N_s/2,N_s\}$ for liquid and vapour branches, and $T_r=T/T_c$ is the reduced temperature.

\begin{figure}
\resizebox{0.5\textwidth}{!}{\includegraphics{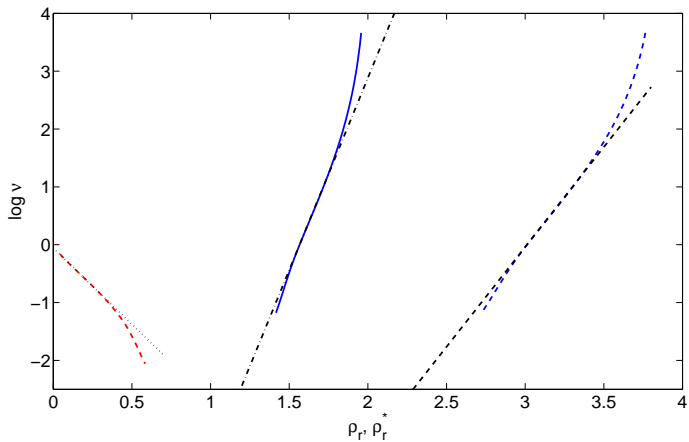}}
\caption{(Colour online) The plot of the reduced inverse fluctuation as a function of the reduced density for the lattice system along a coexistence line. Solid (blue online) and dashed (red online) lines correspond to the liquid and vapour branches. The regions close to the critical point are not shown since there are artifacts emerging due to the finiteness of the system there. Linear fits are presented in Table~\ref{tab}.}
\label{latt_fluct}
\end{figure}

The plot of the inverse reduced fluctuation along a coexistence line resulted from direct {\sc MATLAB} computation using (\ref{nulatt})--(\ref{mN}) for $N_s=1000$, and  is presented in Fig.~\ref{latt_fluct} in a semilogarithmic form as a function of the reduced single-phase density $\rho_{r}=2\langle N\rangle/N_s$. We present only the regions that correspond to the dense liquid (and the related saturated vapour), since the vicinity of the critical point is out of the present study and is poorly reproduced by finite-size calculations.

Within the region of interest, one can see that the semilogarithmic plot of the reduced inverse fluctuation demonstrates sufficiently long  intervals of linear behaviour for both liquid and vapour branches. It should be pointed out that the absolute values of the slope of the liquid branch line (dash-dotted line in Fig.~\ref{latt_fluct}) have practically doubled in value in comparison with the slope of the vapour line (dotted line). This originates from the fact that the most dense packing of particles in a lattice corresponds exactly to the reduced density $\rho_r=2$, while this quantity for real liquids is bounded from above by the value $\rho_r=3.84$ \cite{Cervera2011}. Rescaling $\rho^*_r=(3.84/2)\rho_r$ for the liquid branch gives rise to the dashed line  in Fig.~\ref{latt_fluct}. Within this rescaling,  both slopes (see the Table~\ref{tab}) correspond reasonably well to their value for real fluids for both phases. 

\begin{figure}
\resizebox{0.5\textwidth}{!}{\includegraphics{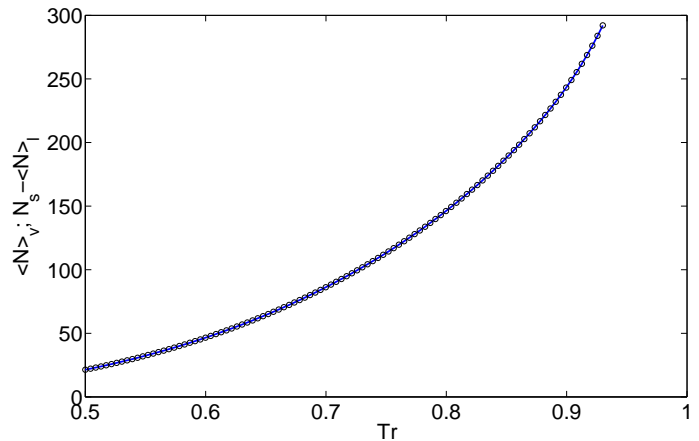}}
\caption{Number of particles (curve)  and vacancies (circles) in the coexisting vapour and liquid phases as a function of the reduced temperature for the simulated lattice fluid system.}
\label{vac}
\end{figure}

To discuss the origin of the approximate equality of the slopes for the semilogarithmic plots of liquid and vapor branches, we have drawn the number of particles belonging to both phases along a saturation line. Fig.~\ref{vac} demonstrates that vacancies in a liquid state and the number of vapour particles are complementary quantities with a high accuracy. Therefore, the model supports the idea that fluctuational properties of  real liquids and the corresponding them vapours along a saturation line, discussed in the previous section,  could  really be explained within the hole theory.

Finally, it is important to note that the vapour branches for both the lattice  (Fig.~\ref{latt_fluct}) and the real fluids (Fig.~\ref{ng_fluct}) demonstrate that this rarefied gas cannot be considered as an ideal gas with respect to structural properties, although they exhibit the PVT properties of ideal gas (and it is usually assumed  in phase coexistence calculations). 
The exponential growth of fluctuations $\langle(\Delta N)^2\rangle\sim \exp(\langle N\rangle)$ (instead of ideal gas $\langle(\Delta N)^2\rangle\sim  \langle N\rangle$) serves as a proof of this.  In fact, only one point $\langle N\rangle=0$ actually corresponds to an ideal gas. Thus, this study of real data on noble gases confirms Rosenfeld's theoretical arguments \cite{Rosenfeld1999} about the importance of taking into account short range intermolecular interactions for modeling entropic properties, even for very dilute real gases.

\section{Discussion and outlook}

It should be pointed out that the standard approaches deal with dilute gases or dense liquids with the goal of testing much more accurate approximations of their structural characteristics, such as the radial distribution function and PVT data. Correspondingly, one adopts different techniques specifically  adjusted for each kind of systems (say, methods of kinetic theory for gases \cite{Chapman1991}, or perturbation methods for liquids \cite{Zhou2011}). 

At the same time, there exists the simple idea of the hole theory of liquids, which describes their structure as comprising a continuous medium with vacancies. In addition, it relates to the Cohen-Turnbull theory of self-diffusion in liquids \cite{Cohen1959}, which predicts an exponential dependence of the self-diffusion on the density, arguing that molecular motion in a dense liquid can be  reduced to a rearrangement of the holes, which form free volume, without changing the molecular structure in general.

In this work, we present the results of consistent studies of reduced inverse fluctuations in a liquid and a corresponding vapour phase along their coexistence curve. The principal points are summarised below.

We introduced a parameter of inverse reduced fluctuations and showed that it is exactly equal to the parameter corresponding to the reduced self-diffusion coefficient, derived by Skripov from a non-equilibrium thermodynamics point of view \cite{Skripov}. Thus, the results of its extraction from experimental data and model calculations allow us to discuss both fluctuation-structural and self-diffusive issues in the same way.

The experimental data for real fluids were used  in this work, in contrast to a majority of approaches where computer simulations for equivalent molecular dynamics ensembles have been applied, e.g. \cite{Bastea2003} (dense liquid argon) and \cite{Goel2011} (silicate melts). Thus, our conclusion about the existence of a relatively wide (up to $25\%$ of reduced density range for each phase) interval where the dependence of the relative fluctuation on the density has a strong exponential character along a liquid-vapour coexistence curve of simple liquids within the region close to the melting point, is a fact of real experimental nature, not derived from model simulations. 

The simultaneous study of both coexisting phases provides the observation that the slopes of the straight lines fitting the logarithms of the relative fluctuations are contrary in signs, but close in absolute values. These results support the assumption that molecules of saturated vapour are principally arranged in such a way that they could fill holes in the coexisting liquid phase. Note, that this idea was formulated in \cite{Skripov} in the context of a self-diffusion mechanism, however without any confirming experimental or theoretical investigations of the vapour phase.

To check this conclusion, we have calculated the inverse reduced fluctuations for coexisting liquid and vapour phases in a lattice fluid for the mean-field grand canonical ensemble. The presence of intervals of similar exponential dependency on the density for both phases, as well as the simultaneously-found equality of the number of  vapour particles and the number of vacancies in the liquid phase, confirms the aforementioned hole theory assumption on the structure of fluids in the determined region.

Finally, the revealed equivalence between the determination of fluctuation via acoustic and thermophysical methods and the characteristics of self-diffusion provides certain perspectives for experimental investigations of the latter. Particularly, it has been shown recently \cite{Yan2008} that the excess entropy extracted from molecular dynamics of self-diffusion provides  important information on structure transitions.  At the same time, the direct experimental measurement of self-diffusion is a very complicated task. Conversely, measurements of the speed of sound and the heat capacity ratio are well-developed techniques with high accuracy. Thus, a study of the logarithm of the introduced inverse reduced fluctuation ratio effectively operates with real, not simulated, substances and analyses their structure transitions.

\end{document}